\documentclass[preprint,preprintnumbers,amsmath,amssymb]{revtex4}

\usepackage[pdftex]{graphicx}
 
\usepackage{dcolumn}
\usepackage{bm}
 
\begin{document}

\title{Use of ultrasound attenuation spectroscopy to determine the size distribution of clay tactoids in aqueous suspensions}

\author{Samim Ali}
 \email{samim@rri.res.in}
\affiliation{Raman Research Institute, C. V. Raman Avenue, Sadashivanagar, Bangalore 560080, India}
\author{Ranjini Bandyopadhyay}
\email{ranjini@rri.res.in}
\affiliation{Raman Research Institute, C. V. Raman Avenue, Sadashivanagar, Bangalore 560080, India}

\date{\today}

\begin{abstract}
  The dispersion processes of aqueous samples of clay are studied using ultrasound attenuation spectroscopy.  The attenuation spectra that are acquired in the frequency range  $10-100$ MHz  are used to determine the particle size distributions (PSDs) for different concentrations and    ages of the clay suspensions. Our analysis, using equivalent spherical diameter (ESD) for circular discs under Stokes drag  in samples of concentrations greater than 1.5\% w/v, shows that a substantial  fraction of the aggregates in suspension are actually tactoids that are composed of more than one platelet. This is in contrast to the general belief that clay disperses into individual platelets in the concentration range where their suspensions exhibit glassy behavior. We conclude that the incomplete fragmentation  of the clay tactoids arises from the rapid enhancement of the inter-tactoid Coulombic repulsion.
  
\end{abstract}
\maketitle     
\section{Introduction}  
    
    Laponite is a synthetic Hectorite clay whose aqueous suspensions have often been used in the study of the glass transition phenomenon. A single Laponite platelet has a thickness of 1 nm and a diameter of 25-30 nm \cite{Kroon_1998}. Because of the highly anisotropic shape and the large negative charges on the flat surfaces, Laponite clays in aqueous suspensions can exhibit a spontaneous ergodicity-breaking phase transition from  free-flowing liquids to non-equilibrium, kinetically arrested states such as gels and glasses at low concentrations ($0.5\%-4\%$) \cite{Ruzicka_2011}. Due to the non-equilibrium nature of these arrested states, they exhibit aging, with the dynamical and thermodynamical properties of these suspensions evolving as the waiting time (measured after the time of preparation) increases. Aging in these systems occurs by the spatially and temporally correlated  local rearrangements of particles that can overcome local energy barriers, with the system exploring progressively deeper energy wells in phase space \cite{Jabbari_2012_1}. The different arrested phases exhibit viscoelastic properties under different applied stress-strain conditions. On application of stresses above a yield value,   the elastic  and viscous moduli of these soft solids decrease and  the suspensions eventually flow like a fluid. Once the applied stress is removed, the suspension gradually recovers its jammed state \cite {Bonn_2002}. Besides their obvious importance in fundamental research, aqueous suspensions of Laponite also find wide applications in the petroleum, cosmetics, pharmaceutical, coating, and food industries \cite{Laponite, scprod}. \\
      It is believed that a major factor contributing to the time-dependent behavior of aqueous  suspensions is the swelling and subsequent breakup of large Laponite clusters into smaller entities \cite{Joshi_2007}. Due to the importance of Laponite suspensions in various technological processes and their frequent use as model glass-formers, the dispersion kinetics of Laponite clay under different experimental conditions needs to be studied. In most of the studies undertaken previously, clusters of clay particles were assumed to exfoliate into individual platelets when dispersed in water. However, several studies also show that clay particles can exist as rigid aggregates, known as tactoids, that consist of more than one platelet oriented parallely. Many experimental methods e.g. Atomic Force Microscopy (AFM), Transmission Electron Microscopy (TEM), Cryogenic Scanning Electron Microscopy (Cryo-SEM), Small Angle X-ray Scattering (SAXS), Small Angle Neutron Scattering (SANS), Dynamic Light Scattering (DLS) and Transient Electrically-induced Birefringence (TEB) have been employed to characterize the tactoid sizes of clays in aqueous suspensions at different concentrations and under varying experimental conditions \cite{Tompson_1992, Saunders_1999, Rosta_1990, Balnois_2003, Bakk_2002, Nicolai_2000}. However, a systematic study of the particle size distribution (PSD) in aqueous clay suspensions and its dependence on clay volume fraction and suspension age is still lacking. Indeed, DLS has been used intensively for particle sizing, but its application is limited to very dilute suspensions where the single scattering mechanism and the Stokes-Einstein relation are valid. Several DLS studies show sol-glass transition with increase in concentration and age of the clay suspensions \cite{Kroon_1996, Bonn_1999epl, Ruzicka_2004, Jabbari_2012_2}. Intensity auto-correlation data obtained in a DLS experiments is typically analyzed in  terms of a fast ($\tau_{1}$) and a slow relaxation time ($\tau_{2}$) \cite{Bonn_2001}. However, it is very difficult to acquire PSD information from $\tau_{1}$ and $\tau_{2}$ in the glassy phase as the Stokes-Einstein relation is invalid in the soft glasses. \\
    
   The aim of this paper is to investigate the dispersion of Laponite clay in aqueous suspensions of different ages using ultrasound attenuation spectroscopy (UAS). The UAS techniques offer a unique possibility of estimating PSDs for soft solid systems and can be used to analyze non-transparent and even highly structured systems. At the same time, it is a non-destructive technique that uses very low intensity ultrasound. The measurement does not, therefore, affect the sample micro-structures and yields an accurate estimate of average particle sizes and PSDs. Over the last two decades, this technique has been applied in different particulate systems of quartz, rutile, latex, alumina particles and  also in emulsions \cite {Dukhin_book}.
     In this work, we measure ultrasound attenuation spectra to estimate PSDs of clay tactoids in Laponite suspensions in a concentrations range ($1.5\%-4\%$  w/v) where the suspensions spontaneously show  liquid to soft solid transitions with increasing age. Furthermore, AFM data has been acquired to confirm the average platelet sizes of both Laponite and Na-Montmorillonite particles in aqueous suspension. We show here that the dispersion kinetics of both clay suspensions follow the same approximate trends.

   
\section{MATERIALS AND METHODS}

\subsection{Material structure and sample preparation}
  
   Two different smectite clays: Laponite and Na-Montmorillonite are investigated in this study. Both the clays are from phyllosilicate group with $2:1$ structure, in which a octahedral metal sub-layer is sandwiched between  two  tetrahedral silica sub-layers. The three sub-layers together form a single unit crystal layer which grows in lateral direction to form a platelet or sheet structure. Some of the metal ions (magnesium in Laponite and aluminium in Na-Montmorillonite) in the octahedral sub-layer get replaced isomorphically by lower valency ions (lithium and magnesium respectively for the two clays). This results in negative charges on the outer surface without any structural change. The rim of the platelet gains  positive charges due to the protonation  of OH$^{-}$ group below a suspension pH of 10.  In dry powder form, the van der Waals interactions between unit layers lead to the stacking of platelets with intercalated sodium counterions. These stacks of clay platelets are known as tactoids.  
   Laponite is a highly pure variety of synthetic clay with a very reproducible chemical composition. According to the manufacturer, the thickness of a Laponite platelet is approximately 1 nm,  while its  diameter is in the range $25 - 30$ nm. The thickness and diameter ranges of individual Laponite platelets, verified using atomic force microscopy (AFM), are shown in figures S1 and S2 of supporting information. 
   Na-Montmorillonite is a natural clay sourced from volcanic ash. The thickness of Na-Montmorillonite platelet is approximately 1 nm, while its lateral size  may vary from a few nanometers to several micrometers.
   Laponite of RD grade (mass density 2.53 g/cc) is purchased from Southern Clay Products and Na-Montmorillonite (mass density 2.60 g/cc) is procured from Nanocor Inc. Clay powder is heated in an oven at a temperature of $120^{\circ}$C for 24 hours before it is dispersed in highly deionized Milli-Q water under vigorous stirring conditions. The powder is added very slowly and in very small amounts to avoid the formation of large aggregates, following which the suspension is stirred vigorously for 45 minutes using an IKA Turrex drive to ensure a homogeneous distribution of the aggregates.

\subsection{Acoustic spectroscopy: Theory and measurements}
 The propagation of sound waves through a medium is closely related to its rheological and acoustic properties \cite{Rayleigh_vol_2}. The crucial rheological parameters are the bulk elastic modulus $M$ and the dynamic viscosity $\eta$, while the basic acoustic parameters are the attenuation coefficient $\alpha$ and the sound speed $c$. Ultrasound spectroscopy measures the attenuation of acoustic waves of MHz frequency by the medium. In this technique, the elastic and dissipative characteristics of the medium are combined into a single complex parameter $k$,  the `compression complex wave number'.  $\alpha$  and  $c$  are  related to  $k$ as follows \cite{Dukhin_book}:
 
 \begin{equation}\label{eqn:attn}
  \alpha = -Im (k)
 \end{equation}
 
 \begin{equation}\label{eqn:c}
 c = \frac{\omega}{Re(k)}
 \end{equation}
 
  Here, $\omega$ is the angular  frequency of the ultrasound waves and $\alpha$ is closely related to the sizes and distributions of the colloidal particles in the suspension medium. When ultrasound interacts with a colloidal sample, dissipation can occur due to a combination of viscous, thermal, scattering, intrinsic, structural, and  electrokinetic processes. If the ultrasound wavelength $\lambda$ is larger than the particle size $a$, $\alpha$ can be written as a sum of the different loss mechanisms, $\alpha = \alpha_{vis} + \alpha_{th} + \alpha_{sc} + \alpha_{int}$, where, $\alpha_{vis}$, $\alpha_{th}$, $\alpha_{sc}$ and $\alpha_{int}$ are the contributions of the viscous, thermal, scattering and intrinsic mechanisms, respectively. The contributions from structural and electrokinetic losses are typically negligible for particulate systems and have not been included here. Depending on the nature and concentration of the colloidal particles under study, it may be possible to separate the different loss mechanisms in the ultrasound frequency domain. In the frequency range of $1-100$ MHz, the viscous loss mechanism  dominates as the aqueous suspensions used here are characterized by high particle density contrasts and rigidities. As the frequency changes from 1 MHz to 100 MHz, the wavelength of ultrasound in water changes from 1 mm to 15 $\mu$m. Scattering loss is negligible as the wavelength range of ultrasound is much larger than the expected sizes of the particles. The expression for the complex wave number $k$, derived in \cite{Dukhin_book} on the basis of coupled phase model \cite{Harker_Temple_1988,Gibson_Toksoz_1989} by assuming predominantly viscous loss in the long wavelength limit ($ka<<1$) can be written as:\\
 
 \begin{equation}\label{eqn:wavenumber}
   \displaystyle\frac{k^{2}M^{*}}{\omega^{2}\rho_{s}} = \displaystyle\frac{1-\frac{\rho_{p}}{\rho_{s}}\displaystyle\sum^{N}_{i=1} \frac{\phi_{i}}{1-(9j\rho_{m}\Omega_{i}/4s^{2}_{i}\rho_{p})}}{1+(\frac{\rho_{s}}{\rho_{p}}-2)\displaystyle\sum^{N}_{i=1} \frac{\phi_{i}}{1-(9j\rho_{m}\Omega_{i}/4s^{2}_{i}\rho_{p})}} 
 \end{equation}
 
 where, \begin{equation}
  s^{2}_{i} = \frac{a^{2}_{i} \omega \rho_{m}}{2 \eta_{m}} 
  \end{equation}
   \begin{equation}
   \rho_{s} = \rho_{p} \phi + \rho_{m}(1-\phi)
   \end{equation}
 and
  \begin{equation}
    M^{*} =  \frac{\rho_{p}\rho_{m}c^{2}_{p}c^{2}_{m}}{\phi\rho_{p}c^{2}_{p}+(1-\phi)\rho_{m}c^{2}_{m}}
    \end{equation}
 Here, $M^{*}$ is the stress modulus,  $\rho_{p}$ is the density of the particle, $\rho_{m}$ is the density of the medium, $\rho_{s}$ is the density of the suspension, $\phi_{i}$ is the volume fraction of the $i$th species of the particulate phase, $\phi$ is the total volume fraction of the particulate phase, $a_{i}$ is diameter of a particle of the $i$th species, $\eta_{m}$ is the shear viscosity of the medium and $\Omega_{i}$ is the drag coefficient of a particle of size $a_{i}$. The Happel cell model is used to incorporate hydrodynamic interactions between the particles \cite{Dukhin_book, Happel_1958}. When Equations 3-6 are coupled with Equations~\ref{eqn:attn} and ~\ref{eqn:c}, $\alpha$ and $c$ can be measured.

  In a typical experiment, if the incident sound intensity is $I_{0}$ and the attenuated intensity after passing through a sample of length ${x}$ is  $I_{x}$, then Beer-Lambert law  gives an expression for the attenuation coefficient $\alpha$ of the medium \cite{Hay_1985}:
  
 \begin{equation}\label{eqn:bl-law}
   \alpha=\frac{10}{x}log\frac{I_{0}}{I_{x}} 
  \end{equation}
 
  Here $\alpha$ is usually expressed in $dB.cm^{-1}$ and depends on the acoustic frequency $\omega$. This law can be easily verified for the colloidal systems studied in this work (Figure S3 of Supporting Information shows data for a 3\% w/v Laponite suspension). The measurements of attenuation coefficients $\alpha$  at varying frequencies give an attenuation spectrum which can be fitted with the theoretical prediction (Equation~\ref{eqn:wavenumber}) using the supplied values of $\rho_{m}$, $\eta_{m}$, $\rho_{p}$, $c_{p}$ and $\phi$ (details are provided in Supporting Information).

  In this work, the attenuation spectra and sound speed are measured using a DT-1200 acoustic spectrometer from Dispersion Technology Inc. Details of this spectrometer can be found in \cite{Dukhin_book}. For each attenuation measurement, 130 ml of the clay suspension is loaded in a cell containing two identical piezoelectric transducers separated by a gap of 20 mm. The transmitting transducer converts input electrical signals into ultrasound tone bursts of different intensities and frequencies. The ultrasound pulse that is generated propagates through the suspension and interacts with the colloidal particles and the liquid medium (water). This interaction decreases the ultrasound intensity, with the predominant loss mechanisms being the viscous loss of the colloids and the intrinsic loss of water. The intensity amplitude of the pulse, received by the detecting transducer, is converted back into an electric pulse for comparing with the initial input pulse. For each frequency, the signal processor receives data for a minimum of 800 pulses to achieve a signal-to-noise ratio of at least 40 dB. The intrinsic loss contribution is subtracted from the total intensity loss to estimate the attenuation arising from the colloids alone. The sound pulses used here are of low power (250 mW) and are not expected to destroy the microscopic structures of the suspensions. The sound speed $c$ is obtained by measuring the time delay between the emitted and received pulses.

 The attenuation spectra  are measured for different ages, $t_{w}$, of the colloidal suspensions by varying the ultrasound frequency in the range 10 to 99.5 MHz with 18 logarithmic steps. $t_{w}=0$ is defined as the time at which the stirring of the sample is stopped. The sample is kept undisturbed during the whole experimental time to prevent destruction of the jammed structures that form as the suspension ages. Prior to these experiments, the sample cell is calibrated for acoustic diffraction using Milli-Q water. All the experiments reported here are performed at 25 $^\circ$C.
 
  An analysis algorithm, developed by the manufacturer and based on the theory described earlier, is used to determine the PSD from attenuation spectra. The algorithm, assuming unimodal or bimodal distributions, runs a search in the size range from 1 nm to 1 mm for the PSD that best fits the data. For a unimodal (lognormal) distribution, the median and the standard deviation of the particle sizes are the fitting parameters.  For a bimodal PSD, essentially a sum of two lognormal PSDs,  the algorithm adjusts four parameters: the two median values ($d_{1}$ and $d_{2}$) of the lower and higher modes,  their standard deviation (assumed to be of the same magnitude $\sigma$ for both modes) and the relative volume fraction ($\phi_{2}$) of particles in the higher mode. During the search process, the algorithm calculates the theoretical attenuation values for each PSD using Equations~\ref{eqn:attn} and ~\ref{eqn:c} and fits the experimental attenuation spectrum by minimizing the fitting error. The searching algorithm stops when the fitting error saturates to the lowest value. The algorithm  performs very well for many standard spherical particles e.g., Ludox TM50 \cite{Dukhin_book} and Silica Koestrosol 1530 \cite{Dukhin_Koestrosol}.  In this work, acoustic measurements are performed on Laponite clay suspensions in a concentration range between 1.5\% w/v and 4\% w/v, and in 1\% w/v Na-Montmorillonite suspensions.

\subsection{Rheology}
 Rheological measurements are carried out at 25$^\circ$C with a stress-controlled rheometer (Anton Paar MCR 501) using a Couette geometry. The sample preparation protocol for rheological measurements are the same as that for the acoustic measurements. For each test, a sample volume of 4.7 ml is loaded in the sample cell just after preparation and is pre-sheared using a high shear stress (60 Pa at an angular frequency $\omega$ of 10 rad/s) for one minute to achieve a reproducible starting point for the aging experiments. The aging process is monitored by recording the evolution of the elastic modulus $G'$ and the viscous modulus $G''$ with aging time $t_{w}$ when an oscillatory strain of amplitude $\gamma_{0}= 0.5\%$ and angular frequency $\omega=1$ rad/s is applied. 

\subsection{AFM measurements}
  A drop of 10 $\mu$l sample is deposited on a freshly cleaved mica surface (area of 1 cm$^2$) and allowed to dry in a closed petri dish under ambient conditions. The AFM micrographs are obtained using a PicoPlus scanning probe microscope (Molecular Imaging Company) operated in tapping or contact mode. In the tapping mode, a cantilever of dimension 225 $\mu$m $\times$ 38 $\mu$m $\times$ 7 $\mu$m with a tip height of 10 $\mu$m, oscillates at its resonant frequency, which in air has a value $146-236$ kHz. In the  contact mode, a cantilever of dimension 450 $\mu$m $\times$ 50 $\mu$m $\times$ 2 $\mu$m with a tip height of 10 $\mu$m is used. Data acquisition and size analysis are done using PicoScan 5.3.3 software supplied by Molecular Imaging Company.

\section{Results and discussion}

\begin{figure}
\begin{center}
\includegraphics[width=4in]{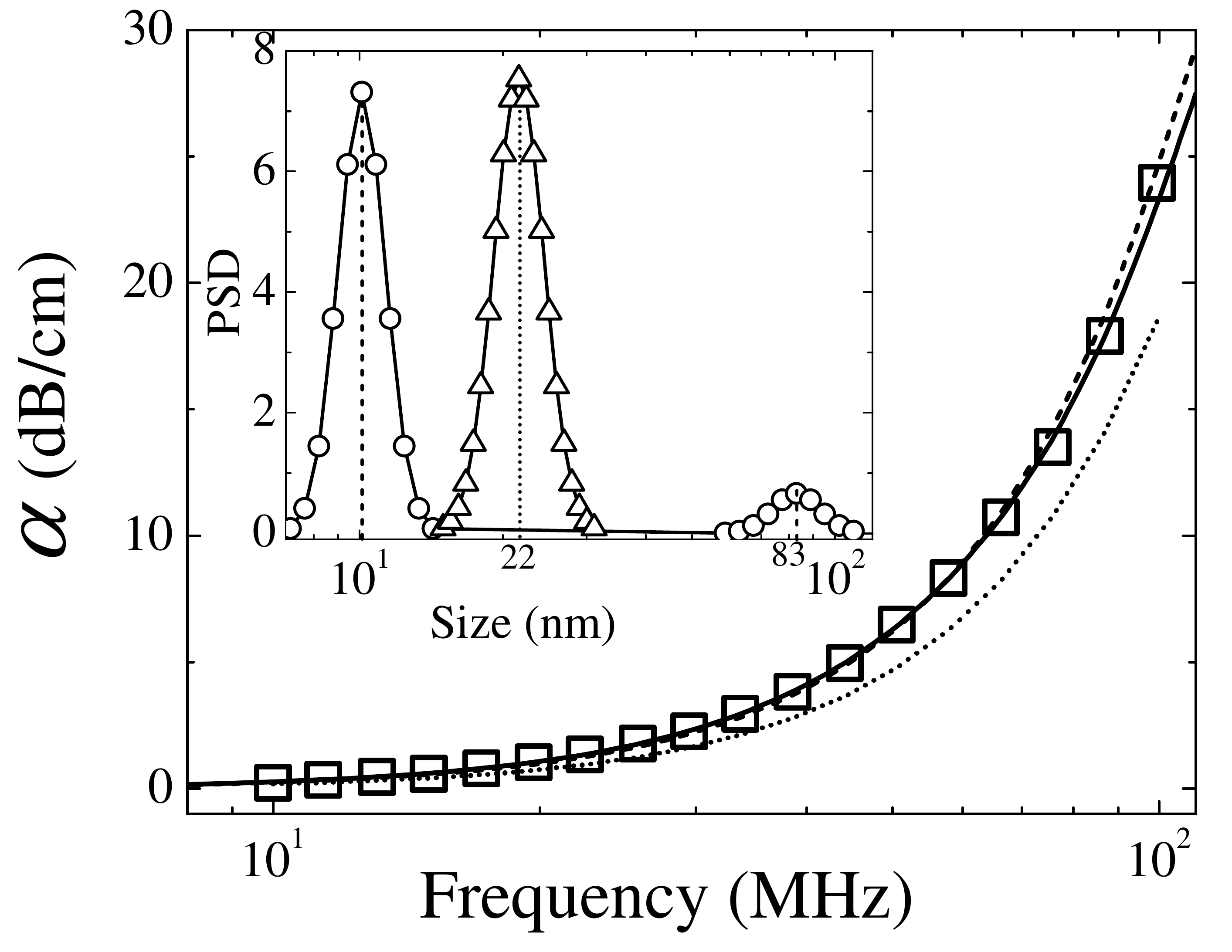} 
\caption{The attenuation spectrum (squares)  for a 3\% w/v Laponite suspension at $t_{w}=6$ hrs. The dashed and solid lines are theoretical fits considering unimodal and bimodal size distributions, respectively. The dotted line corresponds to the intrinsic attenuation in pure water. The unimodal and bimodal particle size distributions (PSDs) obtained from the theoretical fits to the attenuation spectrum are plotted in the inset with triangles and circles respectively.\\}
\label{fig:attn_psd_lapo} 
\end{center}
\end{figure}

  Figure \ref{fig:attn_psd_lapo} shows the measured attenuation spectrum (squares) for a 3\% w/v Laponite suspension of age $t_{w}=6$ hours and the two theoretical fits for unimodal and  bimodal PSDs (dashed and solid lines, respectively). The fitting errors obtained are 3.6\% for the unimodal distribution and 1.3\% for the bimodal distribution. The PSDs obtained from the two fits are plotted in the inset of Figure \ref{fig:attn_psd_lapo} and can be seen to differ substantially.
	
	  As a clay platelet is anisotropic, the viscous loss is dependent on its orientation relative to the sound propagation direction.  A disc moving edgewise causes greater viscous loss than a disc moving broadside \cite{Ahuja_1977, Babick_2006}. The theory discussed above considers the dispersed particles as spheres. The theoretical fits to the experimental data therefore yields an equivalent spherical diameter (ESD) $d_{s}$ which can be related to the platelet dimension using the  Jennings-Parslow relation for anisotropic particle under Stokes drag \cite{jennings_1988}:
\begin{equation}\label{eqn:esd} 
d_{s}= d [\frac{3}{2}\frac{\arctan{\sqrt{(\frac{d}{t})^{2}-1}}}{\sqrt{(\frac{d}{t})^{2}-1}}]^{\frac{1}{2}}
\end{equation}  
Here,  $d$ and $t$ are the diameter and thickness of the disc.\

\begin{figure}
\begin{center}
\includegraphics[width=4in]{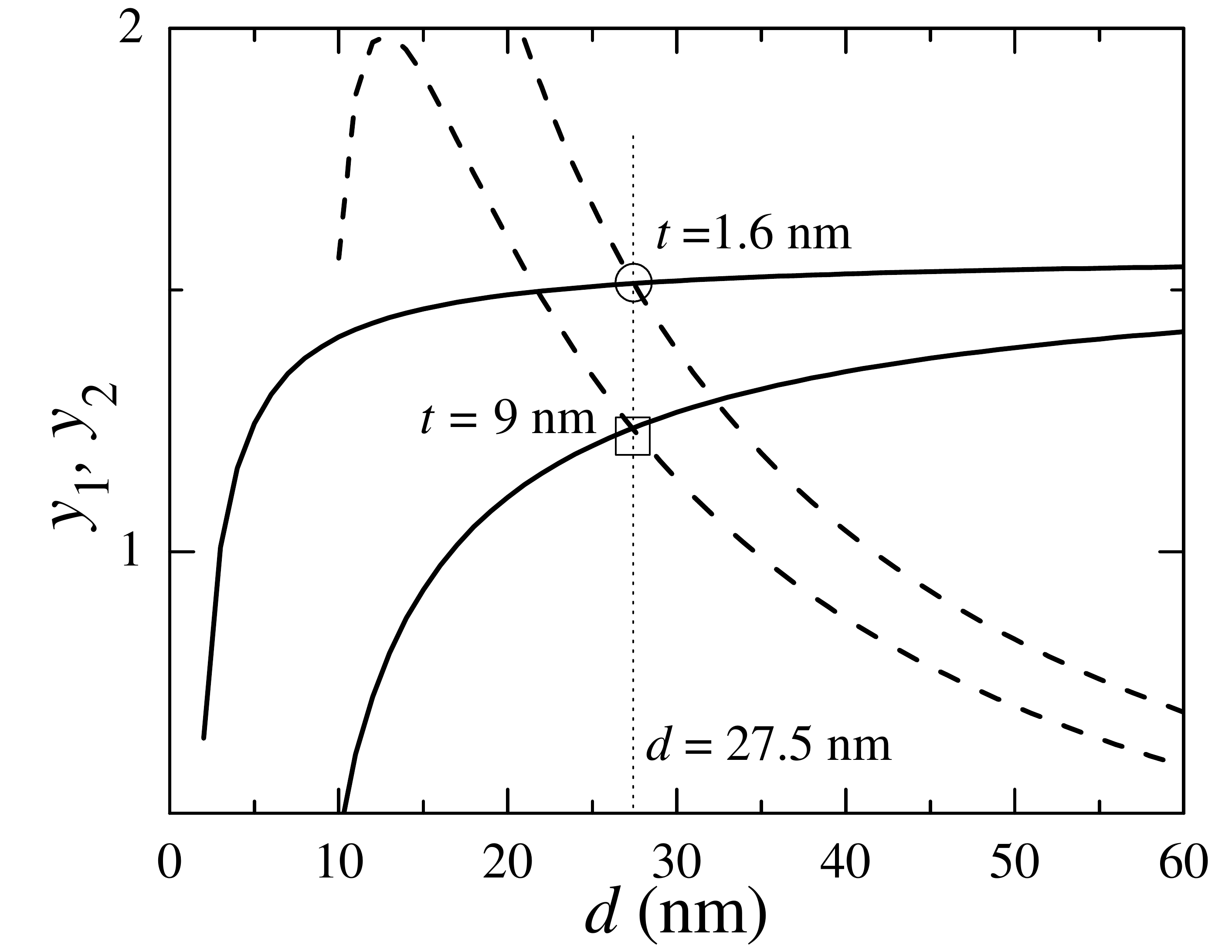}
\caption{ The graphical solutions of Equation~\ref{eqn:esd} for different ESDs. The solution for the median size ($d_{s}=22$ nm) obtained in the unimodal fit corresponds to a Laponite disc of diameter $25-30$ nm when its thicknesses is $t$ = 9 nm (indicated by a square). For the bimodal fit, a median size $d_{1}= 10$ nm is obtained for the lower mode. This corresponds to a disc thickness of $t=1.6$ nm (indicated by a circle). Here, $y_{1}= \arctan{\sqrt{(\frac{d}{t})^{2}-1}}$ is plotted using solid lines and $y_{2}= \frac{2}{3}.\frac{d^{2}_{s}}{d^{2}}.\sqrt{(\frac{d}{t})^{2}-1}$ is plotted using dashed lines.\\}
\label{fig:graphsol_Lapo}
\end{center}
\end{figure}

 The inset of Figure \ref{fig:attn_psd_lapo} shows that the unimodal fit (triangles) to the attenuation curve has a median particle size value of 22 nm. As the basic Laponite platelet is known to be very regular in shape and size, with platelet diameter $25-30$ nm and thickness 1 nm \cite{Avery_1986}, it is possible to estimate the thickness $t$ of a tactoid from  $d_{s}$ by using Equation~\ref{eqn:esd}. Considering a one-dimensional stack of platelets, the optimum graphical solution of Equation~\ref{eqn:esd} (shown in Figure \ref{fig:graphsol_Lapo}) gives a disc diameter $d$ of $25-30$ nm with a thickness $t=9\pm 1$ nm. As the thickness of  a single platelet is 1 nm, this indicates that the majority of the tactoids in suspension are composed of 9 platelets. On the other hand, the bimodal fit (circles in Figure \ref{fig:attn_psd_lapo}) gives two median values: $d_{1}=10$ nm for the lower mode and $d_{2}=83$ nm for the higher mode, with the relative concentration $\phi_{2}$ for the higher mode being around 10\%. From the graphical solution of Equation~\ref{eqn:esd} (shown in Figure \ref{fig:graphsol_Lapo}), $d_{s}=10$ nm, corresponding to the lower mode of the bimodal fit, implies a thickness $t=1.6\pm 0.15$ nm for a disc diameter $d$ of $25-30$ nm. The bimodal assumption is therefore reasonable and confirms earlier reports that claimed the existence of very small Laponite tactoids in suspensions \cite{Tompson_1992, Saunders_1999, Rosta_1990, Balnois_2003}. The spread in the lower mode of the bimodal PSD indicates a size polydispersity and implies that tactoids with more than one platelet are also present in suspension. Further details of the graphical solutions can be found in the Supporting information.

 The contribution to the higher mode of the bimodal distribution (inset of Figure \ref{fig:attn_psd_lapo}), with a median value ($d_{2}$) of 80 nm, is believed to be from larger aggregates and from a very small number of unavoidable bubbles. We should point out here that there is a fair possibility of the incomplete disintegration of some of the clay clusters in the aqueous medium \cite{Bonn_1999}. The presence of small fractions of larger aggregates has also been shown in some previous studies, e.g. in electro-optical experiments on 0.01 wt\%  Laponite suspensions \cite{Zhivkov_2002}.  We therefore believe that modeling the aggregate size distribution with a bimodal function is a better choice than the unimodal function, as the former can efficiently separate the contribution of the big aggregates.

 \begin{figure}
\begin{center}
\includegraphics[width=4in]{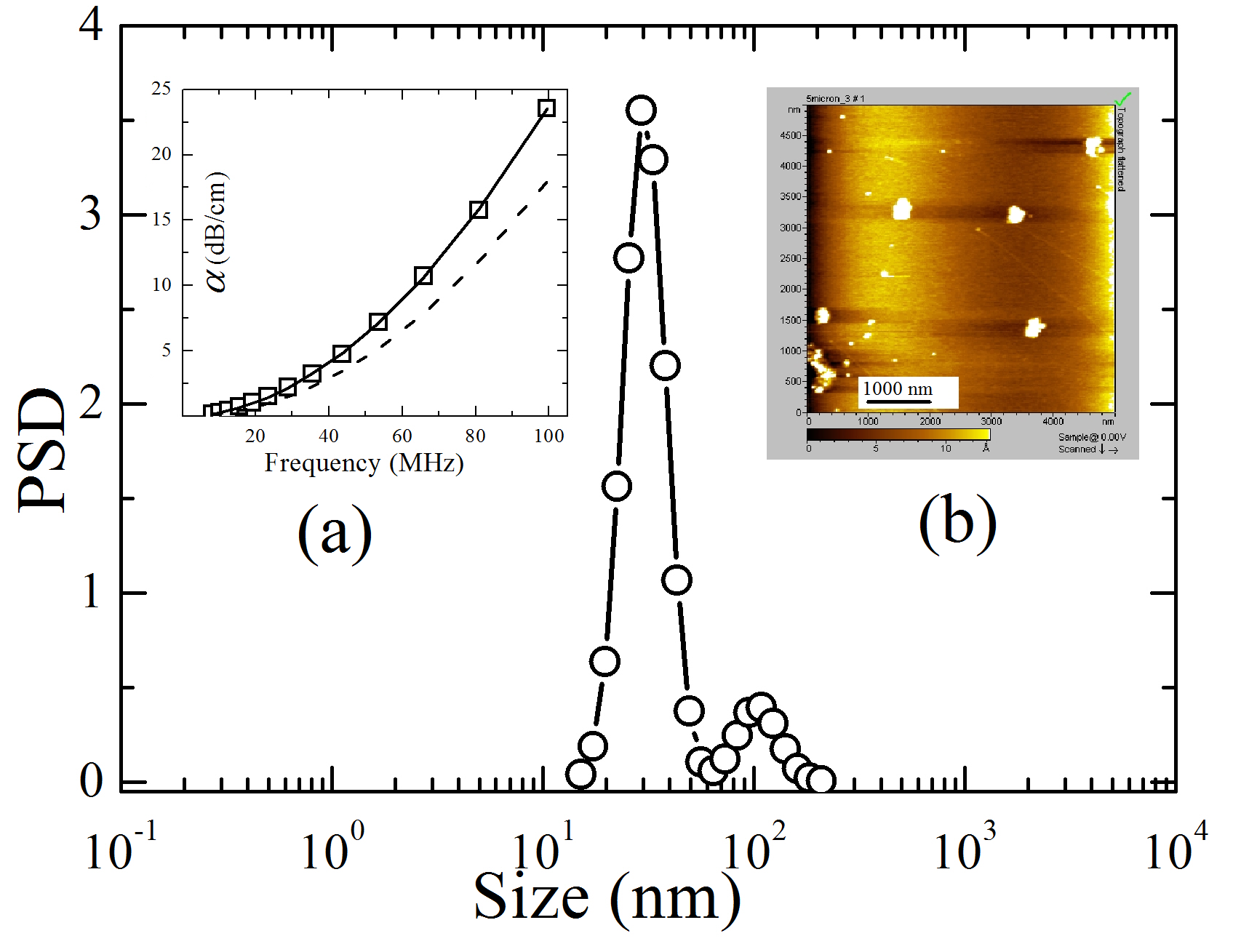}
\caption{The bimodal size distribution of aggregates of a 1\% w/v Na-Montmorillonite suspension at $t_{w} = $17 days. The corresponding attenuation spectrum (attenuation coefficient $\alpha$ vs ultrasound frequency) is denoted by squares in the inset (a). The solid line in the inset (a) is the theoretical fit considering the bimodal distribution, while the dotted line denotes the intrinsic attenuation spectrum in pure water. An AFM image of Na-Montmorillonite sheets (white patches) is shown in inset (b). The scan size is 5 $\mu$m $\times$ 5 $\mu$m.\\}
\label{fig:attn_psd_mmt1}
\end{center}
\end{figure}

   Ultrasound attenuation measurements have also been performed in aqueous suspensions of another anisotropic colloidal clay, Na-Montmorillonite, to verify the techniques already described. Compared to Laponite, Na-Montmorillonite clay particles are more polydisperse in size and have irregular edge boundaries \cite{cadene_2005}. The attenuation spectrum and the aggregate size distribution of a 1\% w/v Na-Montmorillonite suspension of age $t_{w}=17$ days are shown in Figure \ref{fig:attn_psd_mmt1}. As the clay concentration is low and the sample is stirred vigorously for several days, we assume that a substantial part of clay aggregates are exfoliated into  tactoids comprising one to two platelets. Considering a bimodal PSD, the theoretical prediction of the attenuation (shown by the solid line in the inset (a) of Figure 3) fits well to the experimental results (with a fitting error of 1.5\%) to yield $d_{1}=30.1$ nm, $d_{2}=105$ nm and $\phi_{2}=13\%$. The lower mode, $d_{1}=30.1$ nm, gives an equivalent plate diameter $d=254$ nm and a thickness $t=1.5$ nm (graphical solution of Equation~\ref{eqn:esd} is plotted in Figure S4 of Supporting Information). AFM images of the same sample are shown in the inset (b) of Figure \ref{fig:attn_psd_mmt1}. An average size measurement (see Figures S5 and S6 of Supporting Information) of clay sheets yields an average particle diameter $d=282\pm 30$ nm and thickness $t=1.5$ nm  which is very close to the value obtained using the ultrasound measurements discussed above. The median size of the higher mode of the bimodal PSD fit, $d_{2}$, corresponds to  aggregates of several tactoids that exist in suspension because of the incomplete disintegration of the clay powder. These results confirm the feasibility of using ultrasound attenuation spectroscopy  in the study of aging aqueous clay suspensions.\\

In powder form, Laponite clay comprises big clusters of tactoids. Upon dispersing in water, most of the clusters hydrate and disintegrate into smaller entities. The disintegration of the big tactoids happens due to the absorption of water in the successive monolayers, which, in turn, hydrates the sodium ions between a pair of platelets. The osmotic pressure of the hydrated sodium ions pushes the platelets apart in a process that leads to the absorption of more water layers. Finally, the screened Coulombic repulsions that develop between the platelets overcome the intra-platelet van der Waals attractions resulting in further exfoliation.   As a new smaller tactoid is produced, the hydrated sodium ions distribute around the exposed negatively charged surfaces forming diffuse layers that extend into the bulk water phase. The effective volume of each particle increases several times due to the presence of these electrical double layers.\\

\begin{figure}
\begin{center}
\includegraphics[width=4in]{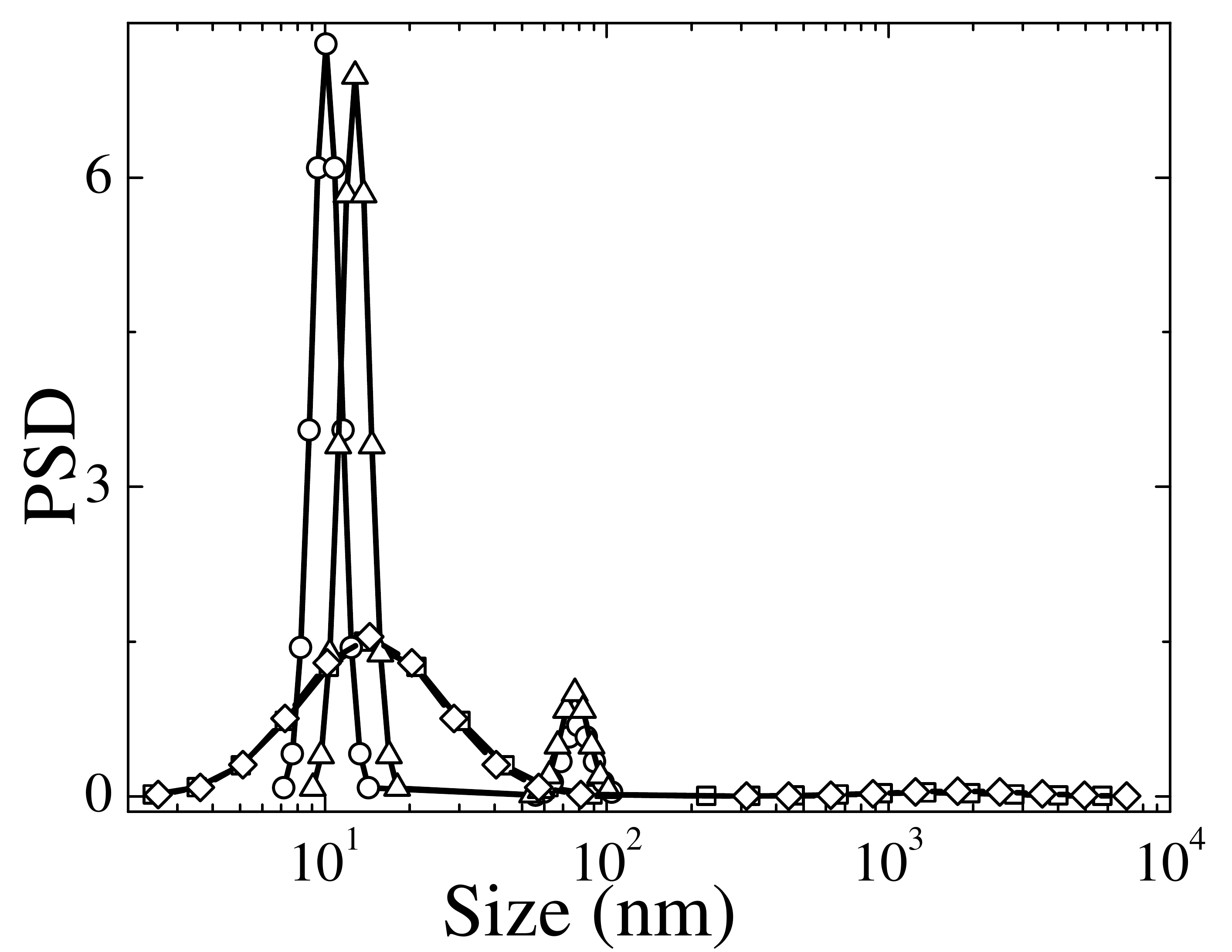}
\caption{Bimodal size distributions of aggregates of 3\% w/v Laponite suspension of $t_{w}=$ 0 hr (diamonds), 3 hrs (squares), 4 hrs (circles) and 58 hrs (triangles).\\}
\label{fig:psd_rd3_age} 
\end{center}
\end{figure}

 It should be noted here that the fragmentation of clay particle is very fast during the initial stages of the stirring process. As the breakup process results in a build-up of strong inter-tactoid repulsions, the subsequent fragmentation becomes slower with time. The evolution of PSDs with $t_{w}$ after the preparation of a 3\% w/v Laponite suspensions is shown in Figure \ref {fig:psd_rd3_age}.  All the PSDs here are obtained from fits to bimodal distributions. For $t_{w}\le 3$ hours, the distributions (data points are denoted by squares and diamonds) show the presence of some big aggregates ($\phi_{2}=4\%$) that measure a few micrometers, while the rest of the aggregates are distributed around a median size $d_{1}$ of 15 nm but with a wide spread in sizes. We estimate the aggregate size in the lower mode by employing graphical solutions of Equation~\ref{eqn:esd} and find that this mode is populated by tactoids composed of three to four tactoids. However, as indicated by the broad distributions of the lower modes of the PSDs, tactoids with more than four platelets could be present. At higher ages, most of the bigger aggregates are fragmented into smaller aggregates and there is the clear emergence of a higher mode of median size $d_{2}=$ 80 nm and $\phi_{2}$= 9\%. The remaining tactoids have median size $d_{1}$ of 10 nm at $t_{w}=4$ hrs (circles in Figure \ref {fig:psd_rd3_age}). At this stage, the lower mode shows a very narrow distribution of sizes and most of the tactoids are composed of one to two tactoids as estimated before. As the sample ages, the size $d_{1}$ in the lower mode increases very slowly. This can be seen from the distribution at $t_{w}=$ 58 hrs (triangles). Since the re-coagulation of clay platelets are prevented due to strong repulsions, the increase in particle sizes  could be due to the slow absorption of one or two layers of water by the tactoids comprising more than one platelet. The complete exfoliation of most of these swollen tactoids are prevented as the inter-tactoid repulsions are much stronger than the intra-tactoid repulsions. A slow growth of the inter-tactoid repulsions has  been indicated in a previous study using X-ray photon correlation spectroscopy \cite{Bandyopadhyaya_2004}. Such an increase in the repulsive force with suspension age explains the ergodic to non-ergodic transition  observed frequently in experiments on Laponite suspensions of concentrations above 2\% w/v \cite{Ruzicka_2011}. The slow exfoliation of tactoids after sample preparation releases intercalated Na$^{+}$ ions into the bulk water. This contributes to an increase in suspension conductivity with age of the clay suspension and has been observed in a previous study \cite{Shahin_2012}.

\begin{figure}
\begin{center}
\includegraphics[width=4in]{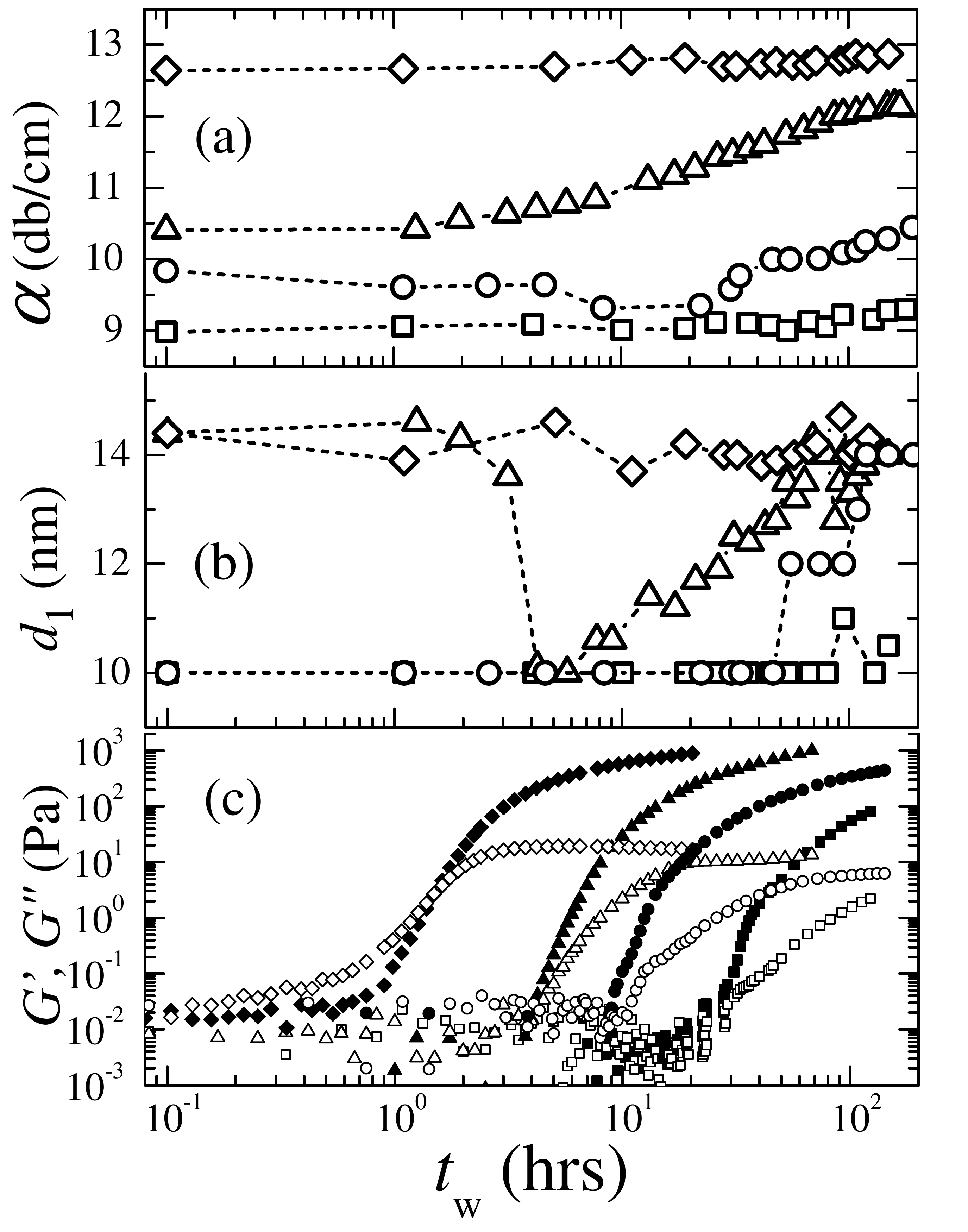}
\caption{Evolutions of (a) the attenuation coefficient $\alpha$ at an ultrasound frequency of 66.3 MHz, (b) the median sizes $d_{1}$ and (c) the elastic moduli $G'$ (solid symbols) and the viscous moduli $G''$ (open symbols) with age $t_{w}$ for 1.5\% w/v (squares), 2\% w/v (circles), 3\% w/v (triangles) and 4\% w/v (diamonds) Laponite suspensions. For rheological measurements, oscillatory shear strains of amplitude $\gamma_{0}=0.5\%$ at an angular frequency of $\omega=1$ rad/s were applied.\\}
\label{fig:attn_d1_moduli-age} 
\end{center}
\end{figure}

 Acoustic measurements are repeated in three other Laponite concentrations: 1.5\%, 2\% and 4\% w/v. The major contribution to ultrasound attenuation is consistently seen  to arise from the particles in the lower modes of the distributions. The evolutions of the attenuation coefficients $\alpha$ at a frequency of 66.3 MHz and the lower median size $d_{1}$ with $t_{w}$ for all the four different concentrations are shown in Figures \ref{fig:attn_d1_moduli-age}(a) and \ref{fig:attn_d1_moduli-age}(b), respectively. Even at other applied frequencies in the range $10 - 99.5$ MHz, the trends reported below are repeated. Due to the slow exfoliation and the swelling of the tactoids described earlier, the attenuation for the 3\% w/v suspension (triangles)  increases slowly over time and shows an approximately 16\% increase before finally saturating. The 2\% w/v sample (circles) shows an increase in attenuation coefficient $\alpha$ and lower mode median size $d_{1}$ at a much later age when compared to the 3\% w/v sample. The other two samples of concentrations 1.5\% w/v (squares) and 4\% w/v (diamonds) do not show any substantial change in $\alpha$, with $d_{1}$ values remaining nearly constant at 10 nm and 14.6 nm, respectively.  As the free volume in the 1.5\% w/v suspension is quite large, most of the clusters break down to single platelets during  the stirring period and a steady state is reached easily. This explains the constant values of   $\alpha$ and $d_{1}$  for the 1.5\% w/v suspension. On the other hand, the clusters in the 4\% w/v suspension disintegrate into tactoids composed of two to three platelets during stirring, following which the suspension rapidly undergoes kinetic arrest due to the highly repulsive interactions between the tactoids. In this case, the inter-tactoid repulsions, which are much higher than the intra-tactoid repulsions,  prevents further swelling and disintegration. This is clearly observed in Figures \ref{fig:attn_d1_moduli-age}(a) and \ref{fig:attn_d1_moduli-age}(b). The enhancement of the inter-tactoid repulsions as a result of the exfoliation of the tactoids is confirmed qualitatively with the age-dependent viscoelastic responses of the Laponite suspensions. The evolutions of elastic modulus $G'$ (solid symbols) and viscous modulus $G''$ (empty symbols) with $t_{w}$ for different concentrations are shown in Figure \ref{fig:attn_d1_moduli-age}(c). Soon after preparation, the 4\% w/v Laponite suspension (diamonds) begins to  show a predominantly elastic response which indicates a jammed state. With aging, both the moduli increase rapidly by several decades due to microscopic rearrangements and the sample finally  exhibits a predominantly elastic response. The 3\% w/v suspension (triangles) starts exhibiting viscoelastic response at an age where the median size of particles drops to a minimum value (Figures \ref{fig:attn_d1_moduli-age}(b) and \ref{fig:attn_d1_moduli-age}(c)). The origin of kinetic arrest in the 1.5 \% w/v Laponite suspension (squares) exhibited at very large ages cannot be related to the repulsion-induced jamming behavior as there is no change in tactoid sizes (squares in Figure \ref{fig:attn_d1_moduli-age}(b)). It has been mentioned earlier that a Laponite platelet contains weak positive charges on its rim \cite{Ruzicka_2011}. The kinetic arrest in 1.5\% w/v suspension can be attributed to the formation of house-of-cards structures (gels) due to the presence of face-rim attractive interactions between the suspended Laponite particles.

\section{Conclusions}
 
          In this work, the exfoliation process of highly anisotropic particles of Laponite and Na-Montmorillonite clays in aqueous suspensions is studied  using ultrasound attenuation spectroscopy (UAS). As the aggregate sizes are smaller than the acoustic wavelength used, the calculated theoretical loss fits well to the experimental data considering only the visco-inertial interactions. The PSDs are extracted by modeling the data using  bimodal distributions. The number of platelets per tactoid is estimated using an ESD formula proposed in \cite{jennings_1988}.  Our analysis confirms the presence of tactoids that consist of more than one platelet in Laponite suspensions of concentrations between 1.5\% and 4\% w/v.  The viscous attenuation for concentrations  below 1.5\% w/v is very small and becomes comparable to the noise of measurement of the intrinsic attenuation. This results in difficulty in data analysis.  Some earlier studies using SAXS, DLS, and AFM have demonstrated the presence of tactoids comprising more than one platelet in aqueous  Laponite suspensions of concentrations less than 1.5\% w/v \cite{Tompson_1992, Saunders_1999, Rosta_1990, Balnois_2003}. A transient electrically induced birefringence (TEB) study on Laponite RD suspensions has also indicated an increase in the average tactoid size with volume fraction of clay particles in the concentration range $0.1-0.8\%$ w/v \cite{Bakk_2002}. However, at very low concentration (0.025\% w/v was studied in \cite{Nicolai_2000}), aggregates are found to quickly disperse into individual discs. The data from these previous studies, along with the results obtained from the ultrasound  spectroscopy experiments reported here, indicate that the tactoid size distributions in aqueous suspensions of clay depend on the concentration of clay particles. The time evolution of PSDs in a 3\% w/v Laponite suspension indicates that the polydispersity of tactoid sizes decreases substantially with age. The same behavior can also be observed in a 3\% w/v aqueous suspension of Na-Montmorillonite (Figure S7 of Supporting Information). The age evolution of the ultrasound  attenuation coefficient $\alpha$ (and hence the PSDs) for different clay concentrations indicates a major role of electrostatic interactions in the tactoid exfoliation process. During the aggregate dispersion process, when the inter-tactoid repulsions becomes comparable to the intra-tactoid repulsions, further exfoliation of the tactoids into smaller entities becomes very slow. We believe that this is the main reason behind the incomplete disintegration of clay clusters  in the concentration range studied here.  Our study  therefore justifies our claim that UAS is a useful technique to elucidate the PSDs of colloidal suspensions whose concentrations lies in a range where other techniques, like DLS or AFM, often fail. 
	
\section{Associated Content}	
 \subsection{Supporting Information}				
  
 AFM images of Laponite tactoids and the estimation of the sizes and the thicknesses of the tactoids are shown in Figures S1 and S2. A plot verifying Beer-Lambert law (Figure S3), supplied values of different constants used in Equation 3, the method of graphical solutions (Figure S4), AFM image analyses of Na-Montmorillonite tactoids (Figures S5 and S6) and a plot for age evolution of aggregate-sizes of Na-Montmorillonite in aqueous suspension (Figure S7) are provided. This material is available free of charge via the Internet at http://pubs.acs.org.

\section{Acknowledgment}
 The authors thank  Dr. A. Dukhin for useful discussions and Mr. A. Dhason for his help in acquiring  AFM images.

\newpage

 \begin{figure}
\begin{center}
\includegraphics[width=6in]{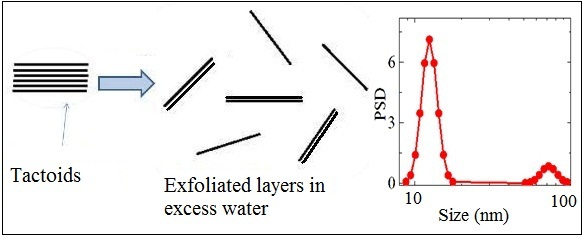}
\caption{\bf Table of contents only}
\end{center}
\end{figure}

\end{document}